\documentclass[a4paper,11pt]{article}
\usepackage{jcappub} 
\usepackage{lineno}
\usepackage{subfig}
\usepackage{booktabs}
\usepackage{multirow}
\usepackage[utf8x]{inputenc} 

\title{\boldmath Multi-messenger particles as a probe for UHECR luminosity}

\author[a, b, 1]{Rodrigo Sasse\note{Corresponding author.},}
\author[c,d]{Adriel G. B. Mocellin}
\author[a, b, c, e]{Rita C. dos Anjos}
\author[b,e]{Carlos H. Coimbra-Araujo}

\affiliation[a]{Programa de p\'os-graduação em F\'isica \& Departamento de F\'isica, Universidade Estadual de Londrina (UEL), 86057-970 Londrina, PR, Brazil}
\affiliation[b]{Applied Physics Graduation Program, Federal University of Latin-American Integration, 85867-670, Foz do Iguaçu, PR, Brazil}
\affiliation[c]{Programa de Pós-Graduação em Física e Astronomia, Universidade Tecnológica Federal do Paran\'a (UTFPR), Av. Sete de Setembro, 3165, 80230-901 Curitiba, PR, Brazil}
\affiliation[d]{Department of Physics, Colorado School of Mines (CSM), 1500 Illinois St, Golden-CO, 80401, USA}
\affiliation[e]{Departamento de Engenharias e Exatas, Universidade Federal do Paran\'a (UFPR), Pioneiro, 2153, 85950-000 Palotina, PR, Brazil}

\emailAdd{rodrigo.sasse1@uel.br}
\emailAdd{agbartzmocellin@mines.edu}
\emailAdd{ritacassia@ufpr.br}
\emailAdd{carlos.coimbra@ufpr.br}

\abstract{Very-high energy (GeV-TeV) gamma rays in the universe suggest the presence of an accelerator in the source. Neutrinos and gamma rays are intriguing astrophysical messengers. Multi-messenger particle emission produced by interactions of cosmic rays with radiation fields and interstellar matter is a probe of luminosity of sources of cosmic rays with EeV energies, known as Ultra-High Energy Cosmic Rays (UHECRs). This work estimates the neutrino flux and suggests that gamma-ray emission is primarily caused by cosmic-ray interactions during propagation. As energy loss processes occur, secondary fluxes are generated, primarily by pion decay. We provide UHECR luminosity of galaxies from multi-messenger particles. These findings not only highlight the potential of certain galaxies as sources of UHECR, but also underscore the intricate interplay of various astrophysical processes within them. By understanding the luminosity patterns and multi-messenger particle emissions, we can gain valuable insights into the environmental conditions, acceleration mechanisms, and other intrinsic properties that position these galaxies as candidates for UHECR production.}

\begin{document}
\maketitle
\flushbottom

\section{Introduction}

The discovery of high-energy neutrinos originating from cosmic sources has provided a novel perspective on studying the Universe. Thus far, the observed arrival directions of the events follow an isotropic distribution, indicating a likely extragalactic source. The current data does not provide any evidence supporting the clustering of events around established astronomical objects, except three potential possibilities, namely TXS 0506+056 \cite{10.1093/mnrasl/sly210}, NGC 1068 \cite{doi:10.1126/science.abg3395} and PKS 1424+240 \cite{doi:10.1126/science.abg3395}. The source BL Lac object TXS 0506+056 was found to be active in high-energy gamma rays and very-high-energy gamma rays with Fermi-LAT and MAGIC telescopes, respectively. PKS 1424+240 is a non-stellar neutrino source, similar to TXS 0506+056, with neutrino excess at the 3.3$\sigma$ level. It can be categorized as a masquerade BL Lac object \cite{Padovani_2022,doi:10.1126/science.abg3395}.  In addition, NGC~1068 -- a nearby type-2 Seyfert galaxy -- was identified with a confidence level of 2.9$\sigma$ as a neutrino source \cite{PhysRevLett.124.051103, murase2019multimessenger} and in 2022 with a global significance of 4.2$\sigma$. In addition, the PKS 1424+240  The emergence of the genesis of high-energy neutrinos has presented itself as a novel enigma within the field of particle astrophysics.

The interaction of high-energy cosmic rays (CRs) with nuclei in astrophysical settings produces neutral and charged pions through the resonant process. These pions then decay, giving rise to the emission of gamma rays and neutrinos (multi-messenger particles). The secondary products, which contain information regarding both the source and the charged particles, can potentially be measured on Earth \cite{ALOISIO201373, 10.1093/ptep/ptx115, ANCHORDOQUI20191}. These particles offer a distinctive means of investigating UHECR sources, as they can traverse the Earth without being influenced by cosmic magnetic fields \cite{Supanitsky_2013,Anjos_2014,Sasse_2021,2021JCAP...10..023D,Coelho_2022}.

With the increasing precision of the measurements of CR fluxes~\cite{AbdulHalim_2023}, we can now accurately evaluate our models of CR propagation in an extensive energy range. In previous research, we used a method to estimate an upper limit on the cosmic-ray luminosity of a single source using the measured upper limit on the integral flux of GeV-TeV gamma rays. Some of the potential sources that are monitored by ground and space-based observatories only provide an upper upper on the gamma-ray flux. The upper limit limits the sum of the primary and secondary photon fluxes for these objects. If the measured upper limit flux of GeV-TeV gamma rays is taken as the upper limit on secondary gamma-rays produced by cosmic rays traveling from a distant source to Earth, the source's cosmic-ray luminosity can be calculated and the model is conservative \cite{Supanitsky_2013, Anjos_2014, Sasse_2021}. In this article, using a new version of CRPropa3, we expand the scope of our investigations to include neutrinos and sources with mixed composition. \cite{2016JCAP...05..038A}. In this respect, the sources NGC 1068, NGC 7755, NGC 175, and Arp 220 are nearby ($D_\text{s} < 80\  \mathrm{Mpc}$) starburst and Seyferts Galaxies capable of producing substantial fluxes of multimessenger particles. Even though these sources are not high-energy neutrino generators \cite{Bechtol_2017, PhysRevLett.116.071101}, they can contribute to emission of charged particles, contributing to the GeV-TeV extragalactic gamma-ray and neutrino fluxes. 

The main contribution of this paper is an extension of the model to extract information on the neutrino physics of the charged nuclei propagating from NGC 1068. Additionally, we present the UHECR luminosity derived from gamma-ray propagation for three sources (NGC 7755, NGC 175, and Arp 220) using a well-established method, providing insights into the cosmic ray physics of these sources and shedding some light on the UHECR source features. The paper is structured as follows: Section 2 introduces the method employed to determine UHECR luminosity through the secondary production of gamma rays and neutrinos resulting from hadronic collisions. Section 3 showcases the results on the galaxies NGC 1068, NGC 7755, NGC 175, and Arp 220. The paper concludes with a comprehensive summary and a discussion of our findings.

\section{Description of the numerical environment from multi-messenger particles}\label{model}

This section will detail the method, which is based on~\cite{Supanitsky_2013,Anjos_2014}. From measurements of the upper limit of the integral of the gamma-ray flux (GeV-TeV) for an individual source, the UHECR luminosity upper limit can be determined. In this study, we focus on those galaxies observed at GeV-TeV energies with an upper limit on the integral of the gamma-ray flux. When direct observations are available instead of only upper limits, the flux includes both source-emitted and propagation-resulting gamma rays. Under such circumstances, we cannot guarantee that the method is conservative and that the UHECR luminosity calculated is total. This method can be applied to any upper limit on the GeV-TeV gamma-ray flux measured by ground or space-based instruments.

\begin{figure}[htbp]
\centering
\includegraphics[width=0.9\textwidth]{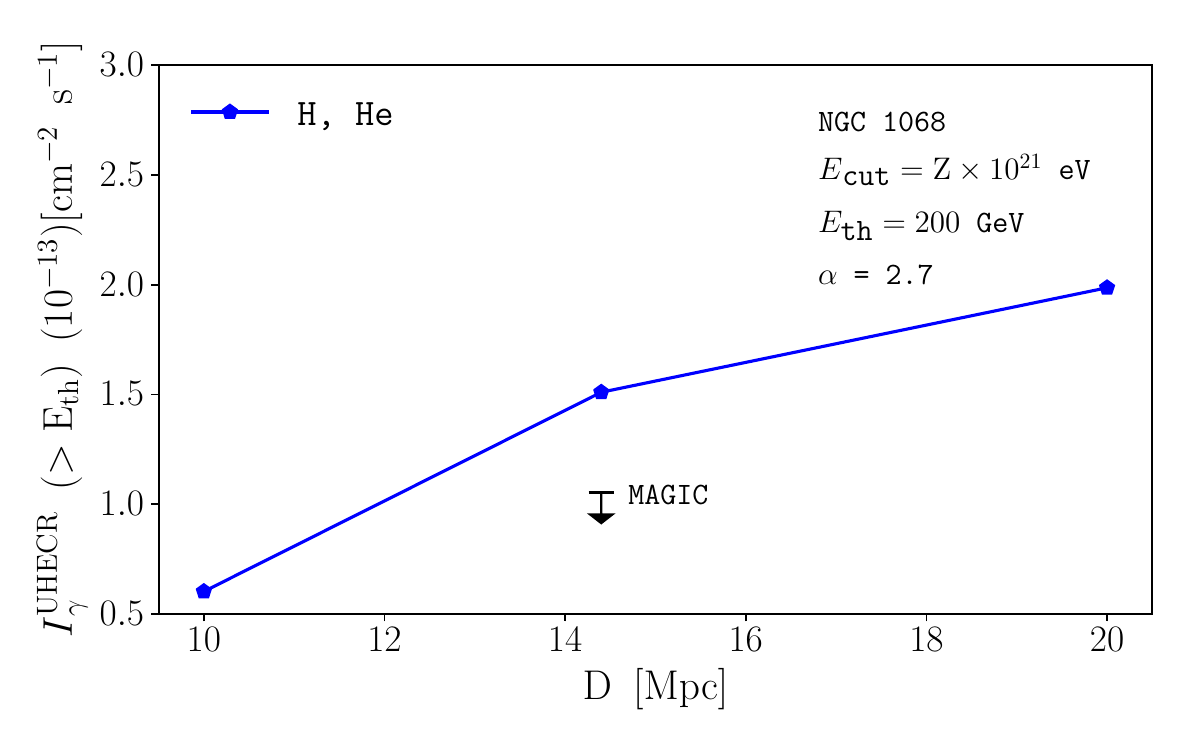}
\caption{Upper limit on the integral flux of gamma rays ($\mathrm{I_{\gamma}^{UHECR}}$) as a function of the source distance obtained using the upper limit of the flux by the Pierre Auger Observatory at 95\% CL for fixed $\alpha = 2.7$, $E_{\mathrm{cut}} =Z\times 10^{21}$ eV, mixed composition at source and $E_{\mathrm{th}} = 200$ GeV. The arrow is the measured
gamma-ray data from MAGIC \cite{2019ApJ...883..135A} (multiplied by the corresponding weight $W_s$).
\label{fig:method}}
\end{figure}

The primary software utilized in this study for conducting simulations of cosmic ray propagation was CRPropa3, a widely adopted tool within the scientific community conducting research on UHECRs. The third version of CRPropa, referred to as CRPropa3 \cite{2016JCAP...05..038A}, includes a variety of modules that enable users to generate a wide range of scenarios regarding the propagation of cosmic radiation. The program is open source, granting users full access to read and modify its codebase. Furthermore, it receives consistent updates, enhancing its capabilities with new tools and improved propagation models\footnote{https://crpropa.desy.de/}. It is assumed that a power-law spectrum per unit solid angle in energy, describes the initial injected cosmic-ray spectrum, as follows: 
\begin{equation}\label{eq:flux}
    \frac{dN}{dE\,dt\,d\Omega} = \frac{L_\text{CR}}{C_0} \hspace{0.1cm}\mathrm{E^{-\alpha}} \hspace{0.1cm}e^{-E/E_\mathrm{cut}},
\end{equation}
where $L_{CR}$ is the cosmic ray luminosity, $C_0 = \int_{E_{\mathrm{min}}}^{\infty} dE E^{-\alpha +1}e^{-E/E_\mathrm{cut}}$ is a normalization constant, $\alpha$ is the spectral index that varies from 2.7 to 3.0, and $E_{cut}$ is the maximum propagation energy. The minimal energy of a cosmic-ray particle produced by the source is $E_{\mathrm{min}} = 10^{18}$ eV.
The cutoff energy is determined as $E_{\mathrm{cut}}^{Z} = Z \times E_{\mathrm{cut}}^{H}$, where Z is the nucleus charge. The spectrum of particle injection at the source is defined by taking into account the energy distribution of particles leaving the source:
\begin{equation}
    \frac{dN}{dE\,dt\,d\Omega} = \frac{L_{CR}}{\langle E \rangle_{0}} \hspace{0.1cm} P_{CR}^{0}(E),
\end{equation}
where $P_{CR}^{0}(E)$ is the normalized energy distribution and $\langle E \rangle_{0}$ is the average particle energy, given by:
\begin{equation}
    \langle E \rangle_{0} = \int_{E_{\text{min}}}^{\infty} \hspace{0.1cm}dE \hspace{0.1cm} E \hspace{0.1cm} P_{CR}^{0}(E) = \frac{\int_{E_{\text{min}}}^{\infty} dE \hspace{0.1cm} E^{-\alpha+1}e^{-E/E_\text{cut}}}{\int_{E_{\text{min}}}^{\infty} dE \hspace{0.1cm} E^{-\alpha}e^{-E/E_\text{cut}}}.
\end{equation}
Considering isotropic emission from a source situated at a distance $D_{\text{s}}$ from Earth and assuming uniform distribution and energy losses, the observed cosmic ray flux on Earth ($I_{CR}(E)$) is
\hypertarget{eq2.6}{\begin{equation}
    I_{CR}(E) = \frac{L_{CR} W_{s}(\hat{n})}{4 \pi D_{s}^{2}(1 + z_{s}) \langle E \rangle_{0}} \hspace{0.1cm} K_{CR} \hspace{0.1cm} P_{CR}(E),
\end{equation}}
where $z_{s}$ denotes the redshift of the source, $\langle E \rangle_{0}$ is the mean energy of particles in the source, $P_{CR}(E)$ is the energy distribution of the particles that reach the Earth, and $K_{CR}$ is the fraction number of cosmic rays that reach the Earth, concerning the number of particles injected. We consider the source weight, denoted $W_{s}(\hat{n})$, which varies based on the source position in the sky and with respect to the exposure of the observatory. Due to the energy-loss mechanisms occurring during cosmic particle propagation, the cosmic ray source generates a secondary gamma-ray flux directly linked to the cosmic ray luminosity. Consequently, the observed gamma-ray flux on Earth is expressed as a function of the cosmic ray luminosity,
\begin{equation}
    I_{\gamma}(E_{\gamma}) = \frac{L_{CR}}{4 \pi D_{s}^{2}(1 + z_{s}) \langle E \rangle_{0}} \hspace{0.1cm} K_{\gamma} \hspace{0.1cm} P_{\gamma}(E_{\gamma}).
    \label{eq:gamma}
\end{equation}
Therefore, $K_{\gamma}$ is the number of gamma rays generated by each cosmic-ray injected into the source, $P_{\gamma}(E_{\gamma})$ is the energy distribution of the gamma rays arriving on Earth and $E_{\gamma}$ is the energy of the gamma rays themselves. The photoproduction of pions, pair-production, and electron-positron annihilation are the main processes that produce gamma rays during cosmic-ray propagation ~\cite{ANCHORDOQUI20191}. Figure \ref{fig:method} shows the simulated $\mathrm{I_{\gamma}^{UHECR}}$ as a function of the source distance for NGC 1068 obtained by using the upper limit flux at 95\% CL of the Pierre Auger Observatory, for $E_{\mathrm{cut}} =Z\times 10^{21}$ eV, $\alpha = 2.7$, $E_{\mathrm{th}} = 200$ GeV and for mixed composition with a proton fraction of 90\% and helium of 10\% as primary. The arrow shows the upper limit on the integral gamma-ray flux of NGC 1068 obtained by MAGIC, at 95\% CL \cite{2019ApJ...883..135A} and $E_{\mathrm{th}} = 200$ GeV. Since our upper limit on the integral of the gamma-ray flux is restrictive, the technique for obtaining the UHECR luminosity can be used.

The upper limit on the integral of the gamma-ray flux is now described as an upper limit (UL) on the cosmic-ray luminosity for an individual source. Therefore, the following expression
\hypertarget{eq2.10}{\begin{equation}
    L_{CR}^{UL} = \frac{4 \pi D_{s}^{2}(1 + z) \langle E \rangle_{0}}{\int_{E^{th}_{\gamma}}^{\infty}dE P_{\gamma}(E_{\gamma})}I_{\gamma}^{UL}(> E_{\gamma}^{th}),
    \label{eq:LCR}    
\end{equation}}
is the upper limit on cosmic ray luminosity from GeV-TeV energy gamma-ray measurements, where $I_{\gamma}^{UL}(> E_{\gamma}^{th})$ is the measure of the upper limit of the integral in the gamma-ray flux obtained by the Observatory at a given energy threshold ($E_{\gamma}^{th}$) and a certain confidence level (CL). Thus, the equation \ref{eq:LCR} summarizes the process, providing insights into potential sources of high-energy cosmic rays.~\citep[see][for detailed reviews]{Supanitsky_2013, Anjos_2014, Sasse_2021}.

\subsection{Neutrino emission from interacting
cosmic-ray nuclei for NGC 1068}

\begin{figure}[h!]
\centering
\includegraphics[width=0.925\textwidth]{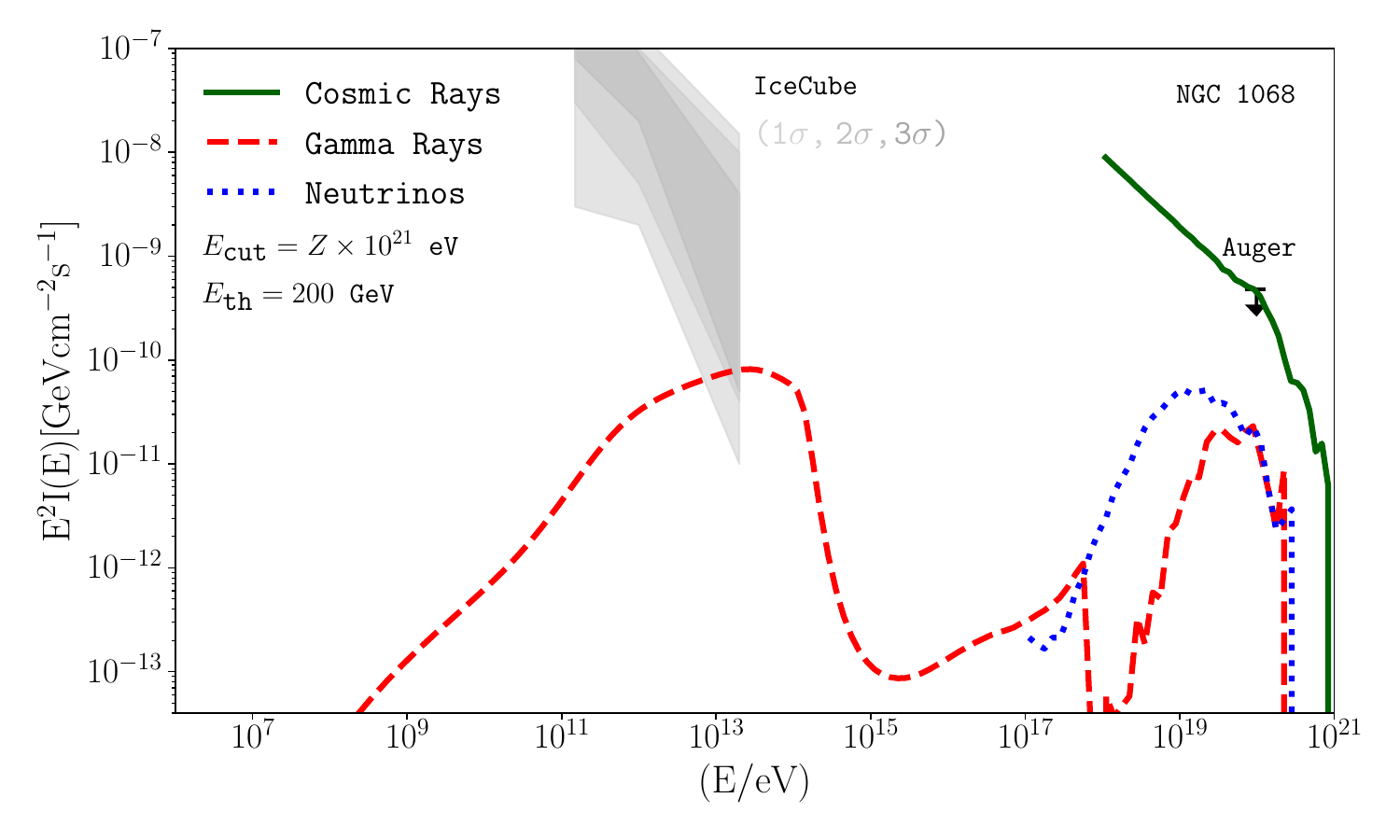}
\caption{High-energy diffuse fluxes of three messenger particles simulated with CRPropa3: CRs and cosmogenic gamma rays and neutrinos for the NGC 1068 according to a power law with an exponential cutoff: spectral index $\gamma = 2.7$ and $E_{cut} = Z \times 10^{21}$ eV. The plot shows the Pierre Auger Observatory spectrum \cite{PhysRevLett.125.121106} and 10 yr IceCube data \cite{PhysRevLett.124.051103}. The incoming cosmic-ray flux was normalized to the flux measured by the Auger Observatory \cite{PhysRevLett.125.121106}. The equivalent normalization was also applied to the gamma-ray and neutrino flux. Primary elements at source were considered to be protons.\label{fig:i}}
\end{figure}

Our aim is to determine a correlation between the neutrino flux originating from the galaxy NGC 1068 and the methodology employed to calculate the upper limit of cosmic ray luminosity. Neutrinos, acting as messenger particles, demonstrate minimal interactions with the intergalactic medium throughout their propagation. Figure \ref{fig:i} illustrates the simulated spectra of UHECRs from NGC 1068 and the secondary particles generated during propagation (gamma rays and neutrinos) by CRPropa3. We adjusted the simulated UHECR spectrum by considering the upper limit determined by the Pierre Auger Observatory at a confidence level of 95\%. The arrow in \ref{fig:i} represents this upper limit used to calibrate the simulated cosmic-ray spectrum. This method allows us to determine the normalization of the secondary GeV-TeV gamma-ray and neutrino spectra. The cosmogenic neutrinos, generated when UHECRs interact with cosmic radiation backgrounds, such as the cosmic microwave background (CMB) and extragalactic background light (EBL), are notably present above $10^{18}$ eV. This scenario aligns with the constraints at exceedingly high energies \cite{2018NatPh..14..396F,2016PhRvL.117x1101A}.

Thus, during the propagation of UHECR particles from NGC 1068, we account for the generation of two secondary fluxes: (I) gamma rays, as described in equation \ref{eq:gamma}, and (II) a secondary flux of neutrinos:

\begin{equation}
    I_{\nu}(E_{\nu}) = \frac{L_{CR}}{4 \pi D_{s}^{2}(1 + z_{s}) \langle E \rangle_{0}} \hspace{0.1cm} K_{\nu} \hspace{0.1cm} P_{\nu}(E_{\nu}).
\end{equation}
Once the maximum UHECR luminosity for a given source is known, it is possible to calculate the maximum neutrino flux to that source. The technique clarifies the paths that lead to the emission of neutrinos and gamma rays, which are secondary particles. This progression plays a crucial role in elucidating the complex dynamics involved, thus providing a more precise perspective that can explain the potential origins of UHECR.

\begin{figure}[htbp]
\centering
\includegraphics[width=0.9\textwidth]{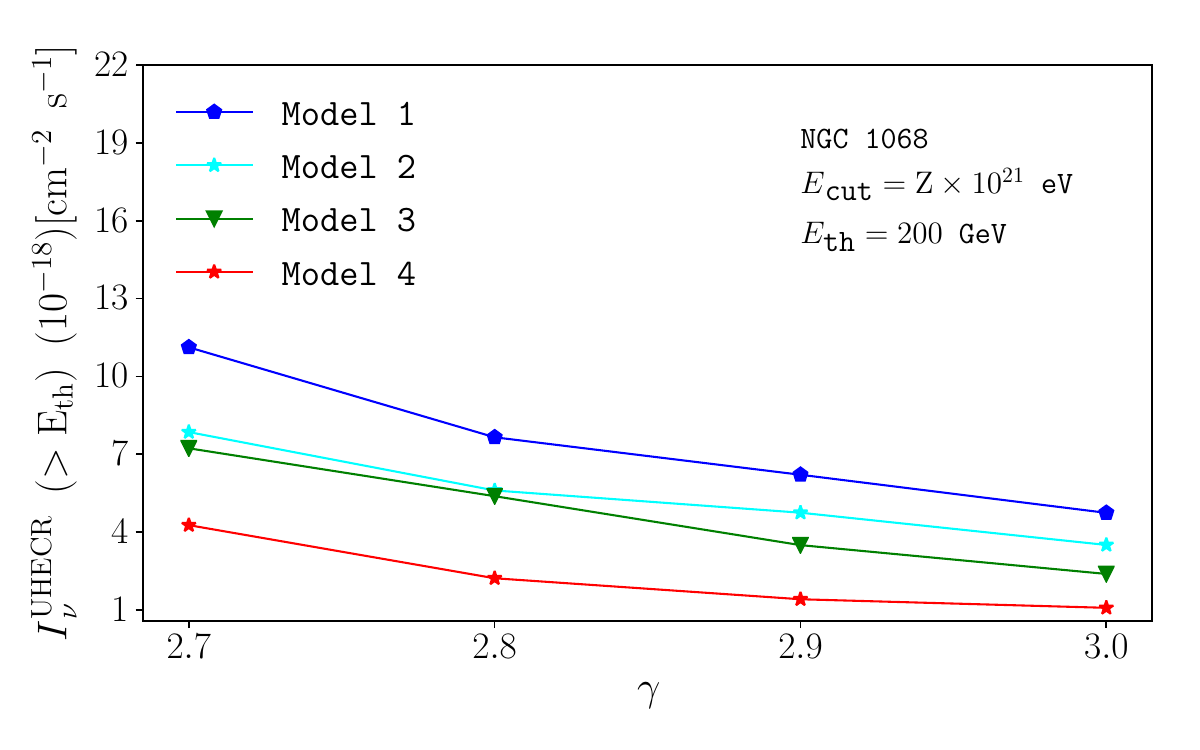}
\caption{Upper limit on the integral flux of neutrinos ($\mathrm{I_{\nu}^{UHECR}}$ at 95\% CL) as a function of the spectral index and for cutoff energy $E_{\mathrm{cut}} =Z\times 10^{21}$ eV.\label{fig:neutrino}}
\end{figure}

\section{Luminosity of cosmic rays sources}\label{environment}

NGC 1068, also known as Messier 77 (\textit{z} = 0.00381), is a spiral galaxy situated in the constellation Cetus. Recent observations by the IceCube Neutrino Observatory revealed an excess in the number of neutrino events originating from the direction of NGC 1068. The observed event count is inconsistent with the expected count, leading the observatory to determine a neutrino flux from that direction, denoted as $ \phi_{\nu} \approx $ 3$\times$10$^{-8}$ (${E_{\nu}}$/TeV)$^{-3.2}$ (GeV cm$^{2}$ s)$^{-1}$ \cite{doi:10.1126/science.abg3395}. Furthermore, the MAGIC collaboration conducted a search for very high-energy gamma rays, conducting 125 hours of observation on NGC 1068. They established an upper limit to the gamma-ray flux at 95$\%$ confidence level, with an energy threshold $E_{\mathrm{th}} = 200$ GeV of $5.1\times 10^{-13}$ $(\mathrm{cm^{2}s})^{-1}$ \cite{2019ApJ...883..135A}. Due to the absence of direct gamma-ray measurements from NGC 1068 by MAGIC at 200 GeV, we relied on the upper limit of the gamma-ray flux derived from the MAGIC Collaboration. Using this information, we can apply our proposed method. By doing so, we determined an upper boundary for the cosmic-ray luminosity of the galaxy. This key step enriched our understanding and provided valuable information on the potential high energy neutrino flux emanating from NGC 1068.

In this scenario, UHECRs undergo electromagnetic acceleration, reaching a maximum energy directly proportional to their electric charge. The spectrum of particles that leave the source environment inside each extragalactic source can be described as a composite of species contributions. The source spectrum is a power-law spectrum with a broken exponential rigidity cutoff, shown in equation (\ref{eq:flux}), with the injection spectra of nuclei: $\alpha: 2.7 - 3.0$, $E_{\mathrm{min}} = 10^{18}$ eV, and $E_{\mathrm{cut}} = Z\times 10^{21}$ eV. For each source distance, simulations were performed for the $10^7$ particles using the one-dimensional approximation.

Figure \ref{fig:neutrino} shows the values of $I_{\nu}^{UHECR}$ for four different fractions of different nuclear mass models described in Tab \ref{tab:1} \cite{2016PhLB..762..288A, 2016arXiv160403637T, 2023EPJWC.28302013G}. \\

\begin{table}[h!]
\centering
\begin{tabular}{c|ccccc}

     Model & &  & $f_{A}(\%)$ &  &  \\

    & H & He & N & Si & Fe \\
\hline
     1 & 90 & 10 & 0 & 0 & 0 \\
     2 & 90 & 5 & 5 & 0 & 0 \\
     3 & 97 & 1 & 1 & 1 & 0 \\
     4 & 96 & 1 & 1 & 1 & 1 \\ 
\end{tabular}
\caption{Mass fractions $f_{A}(\%)$ of various nuclei for models}
\label{tab:1}
\end{table}
The interpretation regarding the physical mass composition is open to discussion, contingent upon the EPOS-LHC, QGSJETII-04, and Sibyll 2.1 hadronic interaction models \cite{de2018testing, 2021Univ....7..321A}. Our approach aligns with the findings of the Auger Collaboration, whose results suggest a shift towards a heavier component around $\sim$ $10^{18.5}$ eV \cite{2014PhRvD..90l2005A}. The spectral index used here enables the determination of an upper limit for the integral of the gamma-ray flux (GeV-TeV) restrictive enough to set an upper limit on the cosmic-ray luminosity \cite{2019ApJ...883..135A}, making our technique applicable \cite{Supanitsky_2013, Anjos_2014}. The upper limit on the integral flux of gamma-rays establishes upper limits on the integral flux of cosmogenic neutrinos for NGC 1068, Fig. \ref{fig:neutrino}. The upper limits on the integral neutrino flux were computed considering cosmogenic neutrinos with energies above $> 10^{17}$ eV. This approach enabled us identify values that are orders of magnitude lower than the gamma values.

The Starburst Galaxy Arp 220 is well-known for intense star formation and is a notable example of the nearest Ultra-Luminous Infrared Galaxy (ULIRG) to us $z \simeq 0.018$. This system results from the merger of two galaxies and encompasses two cores, each approximately 100 pc in radius, with a separation of 350 pc. These cores are enveloped by substantial disks of dense molecular gas, fostering intense star formation. Due to the small size of their cores, an AGN or Starburst phenomenon is necessary to elucidate the remarkably high surface brightness observed in the western core \cite{Sakamoto_2008,Engel_2011}. 

\begin{figure}[h]
  \centering
   \subfloat[Arp 220]{\includegraphics[angle=0,width=0.5\textwidth]{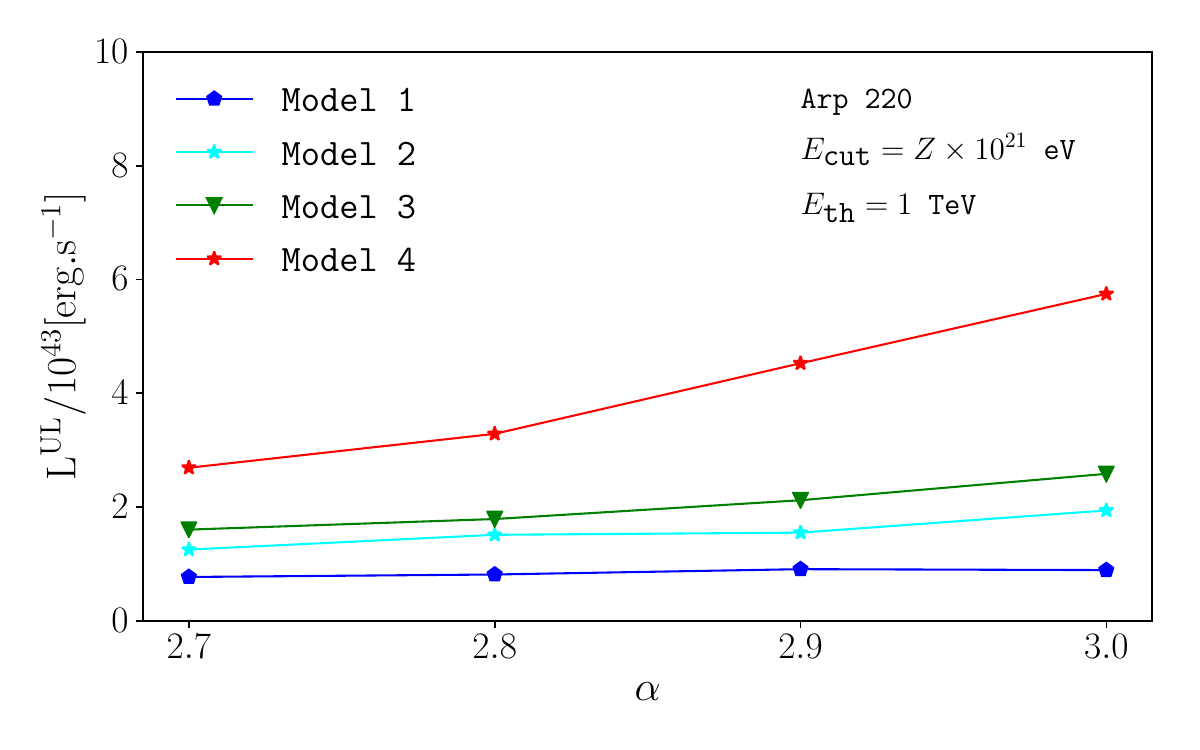}}
    \subfloat[NGC 7755]{\includegraphics[angle=0,width=0.5\textwidth]{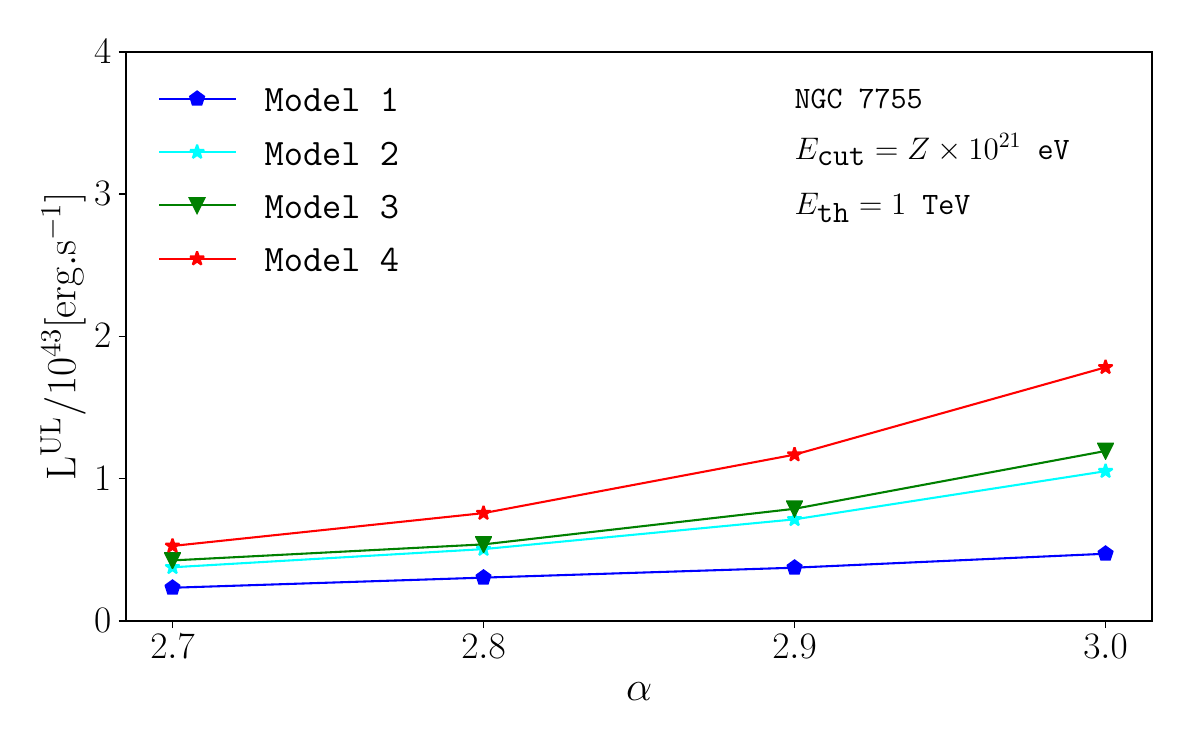}}\\
    \subfloat[NGC 175]{\includegraphics[angle=0,width=0.5\textwidth]{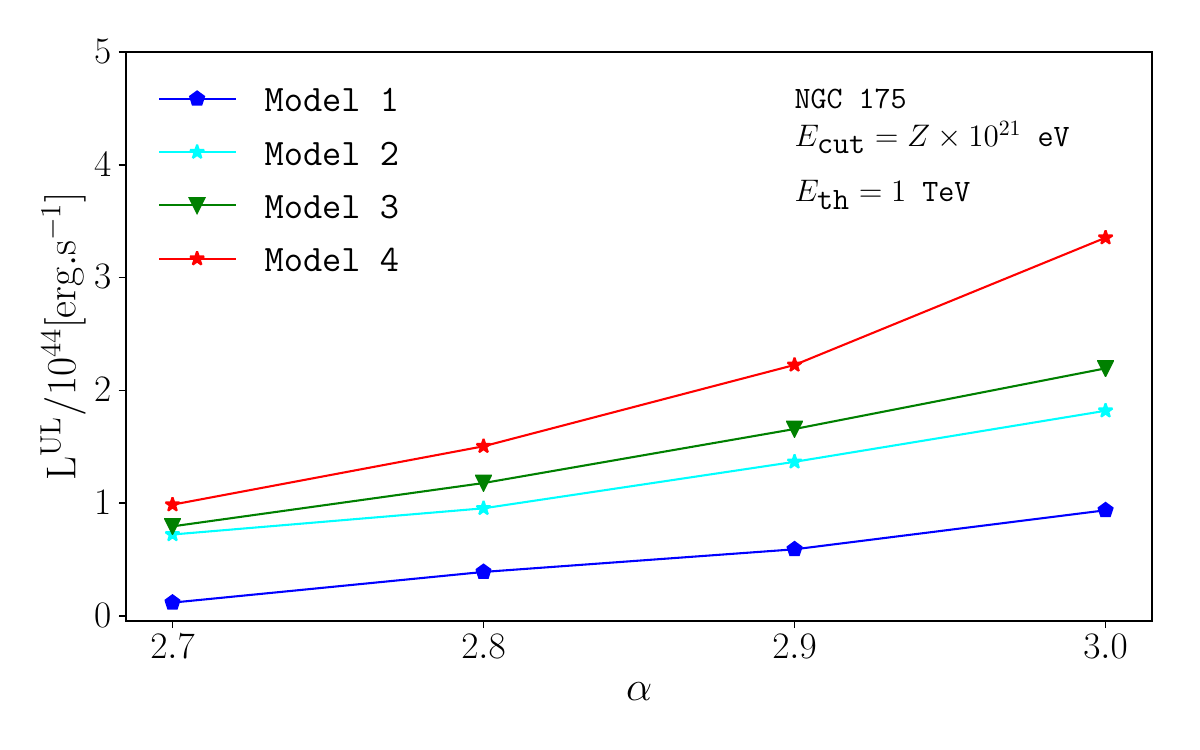}}
    \subfloat[NGC 1068]{\includegraphics[angle=0,width=0.5\textwidth]{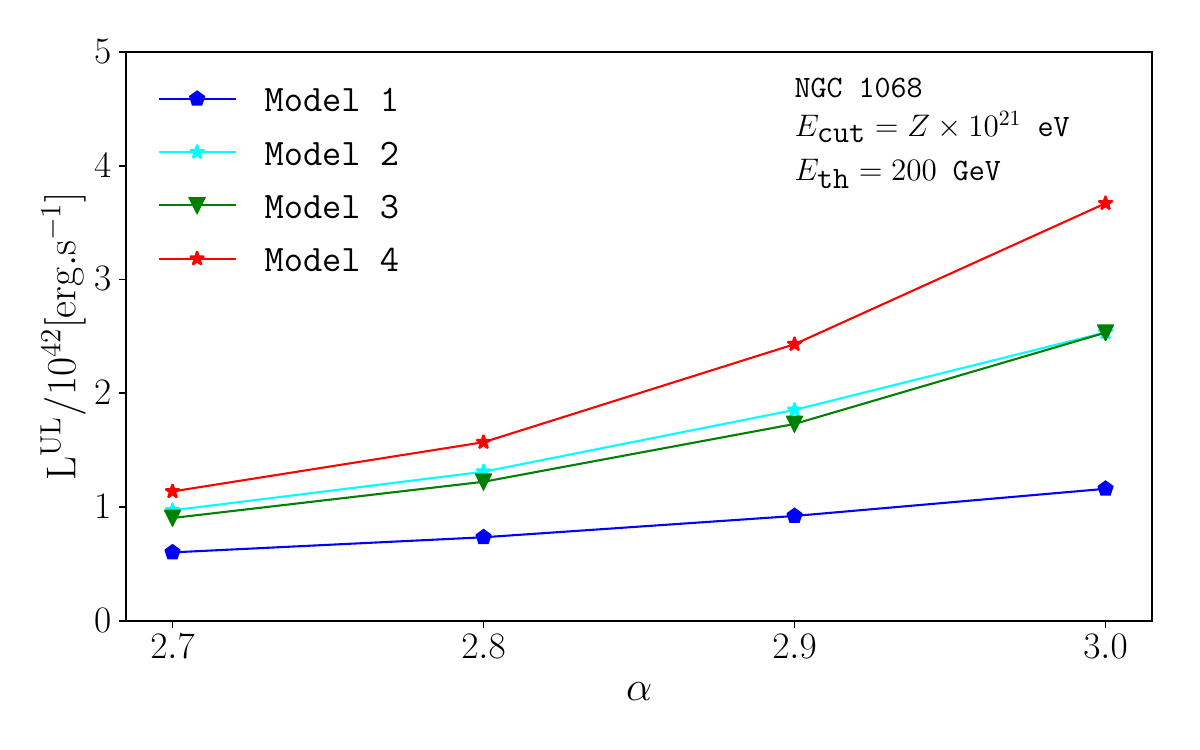}}
   \caption{Upper limits on the mixed cosmic-ray luminosity as inferred from the gamma-ray observations of the sources NGC 7755, NGC 175, Arp 220 and NGC 1068 as a function of the spectral index and for cutoff energy $E_{\mathrm{cut}} =Z\times 10^{21}$ eV.}
    \label{fig:upper}
\end{figure}

Although Arp 220 is categorized as a luminous Starburst Galaxy (SBG), recent investigations suggest that its surroundings do not contribute to the acceleration of the observed UHECRs \cite{Muzio_2023}. Figure \ref{fig:upper}(a) provides the results of the upper limit for UHECR luminosity for Arp 220, taking into account different compositions at the source and spectral index. The limit on the integral gamma of our method was found to be more restrictive in comparison to the limit derived from gamma measurements by the H.E.S.S\@. The upper limit on the integral gamma-ray flux of Arp 220 obtained by H.E.S.S\@, at 99.9\% CL, is $I_{\gamma}^{UL}(E_{\gamma}^{th} > 1\ \mathrm{TeV}) = 1.32\times 10^{-14}\ \mathrm{cm^{2}\ s^{-1}}$ \cite{RC} whereas the lowest value of the integral gamma obtained for our Model 4 with $\alpha = 2.7$ was $I_{\gamma}^{UHECR}(E_{\gamma}^{th} > 1\ \mathrm{TeV}) = 3.39\times 10^{-14}\ \mathrm{cm^{2}\ s^{-1}}$, being more restrictive than the H.E.S.S\@ limit.

For the Seyferts NGC 175 and NGC 7755 hosting supernova remnants (SN), significant gamma-ray emissions have not been observed for any of the objects. Therefore, upper limits on the $> 1$ TeV gamma-ray flux of the order of $10^{-13}\ \mathrm{cm^{2}\ s^{-1}}$ were obtained assuming the confidence level of 95\% \cite{2019A&A...626A..57H}. Supernovae (SNe) emit TeV gamma rays at different stages of their evolution. In addition, core-collapse (cc-)SNe stemming from stellar progenitors can create a rich environment for UHECR acceleration \cite{2019A&A...626A..57H}. The upper limit on the integral gamma-ray flux of NGC 175 hosting SN 2009hf obtained by H.E.S.S\@, at 95\% CL, is $I_{\gamma}^{UL}(E_{\gamma}^{th} > 1\ \mathrm{TeV}) = 5.3\times 10^{-13}\ \mathrm{cm^{2}\ s^{-1}}$ and for NGC 7755 hosting SN 2004cx is $I_{\gamma}^{UL}(E_{\gamma}^{th} > 1\ \mathrm{TeV}) = 1.9\times 10^{-13}\ \mathrm{cm^{2}\ s^{-1}}$. Our lowest restrictive upper limits on the integral gamma-ray flux for our Model 4 with $\alpha = 2.7$ were $I_{\gamma}^{UHECR}(E_{\gamma}^{th} > 1\ \mathrm{TeV}) = 5.59\times 10^{-13}\ \mathrm{cm^{2}\ s^{-1}}$ and $I_{\gamma}^{UHECR}(E_{\gamma}^{th} > 1\ \mathrm{TeV}) = 2.09\times 10^{-13}\ \mathrm{cm^{2}\ s^{-1}}$ for NGC 175 and NGC 7755, respectively. The plots in Fig. \ref{fig:upper}(b) and (c) show the upper limits for UHECR luminosity concerning the hosts of SNe SN 2009hf and SN 2004cx in Seyfert sources, utilizing values obtained from a more restrictive upper limit on the integral gamma-ray.

\section{Discussion and concluding remarks}\label{results}

In this study, we derived the upper limits on cosmic ray luminosity for four distinct sources by analysis multi-messenger: supernovae within three Seyferts-type and Starburst galaxies. The upper limit measurements of gamma-ray flux were obtained using the H.E.S.S and MAGIC Observatories. For every source analyzed in our study, we determined the maximum UHECR luminosity by integrating these measurements with our method. These results emphasize the importance of taking into account the intricate dynamics of cosmic-ray propagation from their originating sources to Earth and the nuanced interplay of factors governing such processes.

\acknowledgments

We thank the anonymous referee for the pertinent ideas and suggestions that significantly enhanced this work. This study was financed in part by the Coordenação de Aperfeiçoamento de Pessoal de Nível Superior – Brasil (CAPES) – Finance Code 001. Research by RCA is supported by Conselho Nacional de Desenvolvimento Cient\'{i}fico e Tecnol\'{o}gico (CNPq) grant numbers 307750/2017-5 and 401634/2018-3, and Serrapilheira Institute grant number Serra-1708-15022. She also thanks for the support of L'Oreal Brazil, with partnership of ABC and UNESCO in Brazil. The authors acknowledge the National Laboratory for Scientific Computing (LNCC/MCTI, Brazil) for providing HPC resources of the SDumont supercomputer, which have contributed to the research results reported in this paper. URL: http://sdumont.lncc.br. R.C.A. also acknowledge FAPESP Project No. 2015/15897-1. R.C.A. and C.H.C.-A. acknowledge the financial support from the NAPI “Fenômenos Extremos do Universo” of Fundação de Apoio à Ciência, Tecnologia e Inovação do Paraná.

\bibliographystyle{JHEP}
\bibliography{biblio.bib}

\end{document}